\documentclass[aps,prl,twocolumn,superscriptaddress,showpacs,floatfix,amsmath,amssymb]{revtex4-1}
\usepackage{graphicx} 
\usepackage{bm}
\usepackage{color,mathtools}    
\usepackage{hyperref}    
\usepackage{float}    
\usepackage{mciteplus}
\usepackage{wrapfig}

\def\K{\mathbf{K}}
\def\k{\mathbf{k}}
\def\tildek{\mathbf{\tilde k}}

\begin{document}

\title{First-principles-based method for electron localization: Application to monolayer hexagonal boron nitride}

\author{C. E. Ekuma}
\altaffiliation{National Research Council Research Associate}
\email{chinedu.ekuma.ctr.ni@nrl.navy.mil}
\affiliation{Naval Research Laboratory, Washington, District of Columbia 20375, USA}

\author{V. Dobrosavljevi\'{c}} 
\affiliation{Department of Physics and National High Magnetic Field Lab., Florida State University, Tallahassee, FL 32306, USA}

\author{D. Gunlycke}
\email{daniel.gunlycke@nrl.navy.mil}
\affiliation{Naval Research Laboratory, Washington, District of Columbia 20375, USA}


\begin{abstract}    
\noindent We present a first-principles-based many-body typical medium dynamical cluster approximation method for characterizing electron localization in disordered structures.  This method applied to monolayer hexagonal boron nitride shows that the presence of a boron vacancies could turn this wide-gap insulator into a correlated metal.  Depending on the strength of the electron interactions, these calculations suggest that conduction could be obtained at a boron vacancy concentration as low as $1.0\%$.  We also explore the distribution of the local density of states, a fingerprint of spatial variations, which allows localized and delocalized states to be distinguished.  The presented method enables the study of disorder-driven insulator-metal transitions not only in $h$-BN but also in other physical materials.
\end{abstract}    
    
\maketitle     

\begin{figure}[b!]
	\centering
	\includegraphics[trim = 0mm 0mm 0mm 0mm,scale=0.25,keepaspectratio,clip=true]{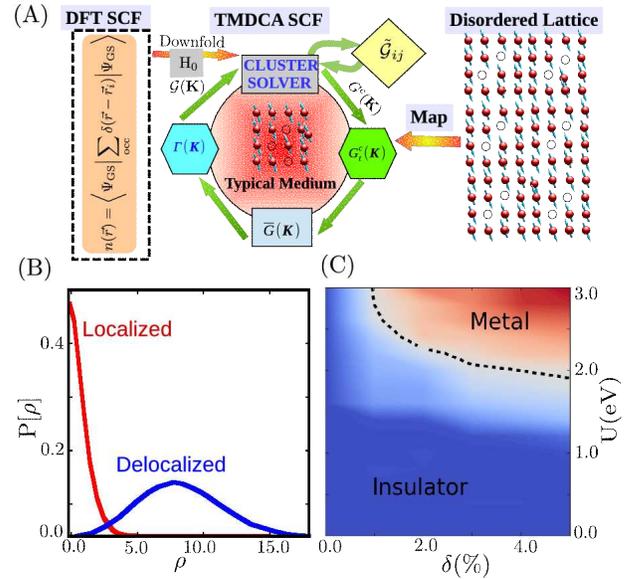}
	\caption{(A) Schematic of the TMDCA@DFT method.  The DFT SCF solution is downfolded and used in the primary TMDCA SCF routine of a typical medium mapped by the disordered lattice.  The cluster solver has also a secondary SCF routine to account for the response of electron interactions [see the Supplemental Material (SM)~\cite{Supp} for details].  (B) Illustration of a local density of states distribution $\mathrm{P}[\rho]$ that is approximately Gaussian (log-normal) for delocalized (localized) states. (C) The typical density of states, defined below, as a function of boron vacancy concentration $\delta$ and Hubbard $U$ calculated at the Fermi level in monolayer hexagonal boron nitride.  The typical density of states per unit cell [Fig.\,\ref{f.2}(A)] ranging from 0 (blue) to $0.065$\,eV$^{-1}$ (red) shows an insulator-metal transition at roughly $\delta\times U^4\approx 0.8$\,eV$^4$. The dashed line is intended to give a rough estimate of the location of the transition in parameter space.}
	\label{f.1}
\end{figure}

Pioneering work on electron localization arising from disorder~\cite{Anderson1958}, electron interactions~\cite{10.1139,*Mott,*Mott1967,*Mott1949}, and a combination of both~\cite{Altshuler,*Efros1975,*Finkel1983a,*PhysRevB.63.172403,Lee1985,*RevModPhys.66.261,50years,*Hirschfeld-review,*efros1985electron,*Miranda2012} have been shown to lead to diverse emerging phenomena in a wide range of physical systems, one of which is the insulator-metal transition (IMT)~\cite{Anthony2010,PhysRevB.71.125104,*Sahu2011523,*Raychaudhuri,*PhysRevB.73.125411,*RevModPhys.84.1067}. Though disorder and electron interactions can independently lead to an IMT, transport and scanning probe measurements have shown that both are needed for a proper characterization of real materials~\cite{Lee1985,*RevModPhys.66.261,Anthony2010,nn103548r}. Computational approaches for studying correlated, disordered materials generally rely on either density functional theory (DFT)~\cite{Hohenberg1964,*Kohn1965} using supercells or the dynamical mean-field approximation (DMFA)~\cite{PhysRevLett.52.77,*PhysRevB.36.1111}, including cluster extensions~\cite{PhysRevB.63.125102,*PhysRevB.61.12739,*Jarrell01,RevModPhys.77.1027}.  While the DFT supercell approach can only describe ordered defect structures, DMFA deals explicitly with statistical disorder distributions~\cite{PhysRev.156.809,*Velicky68}, as does the coherent potential approximation (CPA).  
Traditional DMFA/CPA methods use arithmetic averages in their self-consistent-field (SCF) routines, and therefore lose essential information about the distribution of the local density of states $\rho$.
Moreover, these methods use arithmetically averaged density of states, $\rho_a\equiv\langle\rho\rangle_\mathrm{arit}$, which cannot distinguish between extended and localized states~\cite{Vlad2003,PhysRevB.89.081107}.  Therefore, we will instead adopt the typical medium dynamical cluster approximation (TMDCA)~\cite{Vlad2003,PhysRevB.89.081107,PhysRevB.92.201114}, which is built around a geometrically averaged density of states $\rho_g\equiv\langle\rho\rangle_\mathrm{geom}$.  This latter average is sensitive to skewness in the local density of states distribution $\mathrm{P}[\rho]$~[see Fig.\,\ref{f.1}(B)], making $\rho_g$ a suitable \textit{order parameter} for characterizing localization transitions~\cite{nakayama2003fractal,*Janssen1998,*Thouless1974,*Thouless1970,PhysRevB.89.081107,0953-8984-26-27-274209,Vlad2003}.  The TMDCA approach has both experimental and theoretical support~\cite{Vlad2003,PhysRevB.89.081107,PhysRevB.92.201114,PhysRevB.81.155106,Anthony2010,Li2013,PhysRevB.92.014209,*PhysRevB.90.094208,*PhysRevB.92.205111} and has been successfully used to describe disordered and/or interacting model systems~\cite{PhysRevB.89.081107,PhysRevB.92.201114}.

In this letter, we extend the TMDCA approach to physical materials with electronic properties computed from density-functional theory (TMDCA@DFT).  See schematic in Fig.\,\ref{f.1}(A).  This first-principles-based many-body approach is expected to provide further insight into electron localization and IMTs in real materials.  Herein, we apply the TMDCA@DFT approach to explore a correlation-mediated IMT in monolayer hexagonal boron nitride ($h$-BN)~\cite{doi:10.1021/nl103251m,PhysRevB.80.155425,doi:10.1021/nl9011497} shown in Fig.\,\ref{f.2}(A).  As a two-dimensional (2D) crystal~\cite{Novoselov666,*PhysRevB.91.035402,RevModPhys.81.109,doi:10.1021/nl103251m,doi:10.1021/nl1022139,PhysRevB.80.155425,doi:10.1021/nl9011497}, monolayer $h$-BN is a candidate material for use in electronics~\cite{doi:10.1021/nl103993z}.  The challenge is that unlike graphene, $h$-BN lacks inversion symmetry~\cite{0370-1298-65-10-307,*PhysRevB.79.115442,*doi:1426575} with the difference in electronegativity between B and N sites making $h$-BN a wide-gap insulator~\cite{doi:10.1021/nn1006495}.  It has been shown that strain-engineering could shrink the band gap~\cite{RevModPhys.81.109,PhysRevLett.98.206805,*nl2035749} but so far not enough to make $h$-BN useful for field-effect transistors (FETs).  Motivated in part by the recent experimental observation of IMTs in $h$-BN nanostructures~\cite{nn103548r} and related materials~\cite{4278352,*12892659,*Shukla2015,*doi:10.1021/nl4007479}, we explore the possibility to obtain conduction in $h$-BN through a disorder-induced IMT.  Our calculations reveal an IMT requiring the presence of both electron interactions and disorder with the transition following a monotonic curve [see Fig.\,\ref{f.1}(C)].  Assuming a Hubbard $U$ similar to that in graphene with $U=2.7$\,eV~\cite{PhysRevB.75.064418,*nl0717917,*PhysRevLett.115.186602}, our calculations suggest that the IMT could potentially be found at an averaged boron vacancy concentration $\delta$ as low as $1.0\%$.

\begin{figure}
	\centering
	\includegraphics[trim = 0mm 0mm 0mm 0mm,scale=0.25,keepaspectratio,clip=true]{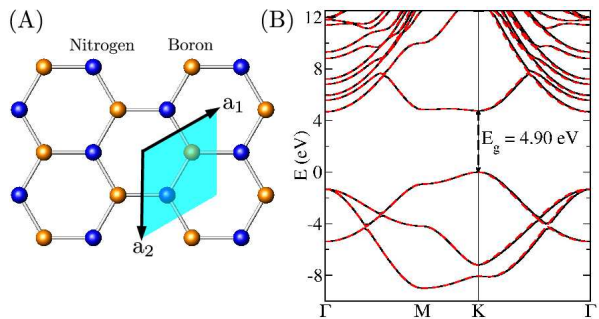}
	\caption{(A) The hexagonal structure of monolayer $h$-BN with a highlighted unit cell defined by the lattice vectors $\vec{a}_1$ and $\vec{a}_2$.  (B) The DFT band structure (solid black bands) reproduced by the downfolded Hamiltonian $H_0$ (red dashed bands).  The large band gap $E_g$ at the Brillouin zone corner points K makes pristine $h$-BN an insulator.}
	\label{f.2}
\end{figure}


To study localization in the presence of disorder and electron interaction, we adopt the Anderson-Hubbard Hamiltonian of the form
\begin{equation}
	H = H_0+\sum_{i\alpha\sigma} V_{i\sigma}^\alpha n_{i\sigma}^\alpha + U\sum_{i\alpha} n^\alpha_{i \uparrow} n^\alpha_{i \downarrow},
	\label{e.1}	
\end{equation} 
where $H_0$ is the single-particle Hamiltonian, $V_{i\sigma}^\alpha$ is a disorder potential, and $n_{i\sigma}^\alpha$ is the number operator, where $i$, $\alpha$, and $\sigma$ are site, orbital, and spin indices, respectively. The three terms above describe respectively single-particle, disorder, and electron interactions.

Herein, we obtain the single-particle Hamiltonian from DFT~\cite{Hohenberg1964,*Kohn1965} calculations using the linearized augmented plane-wave method, as implemented in WIEN2K~\cite{Blaha2001}. Structural and electronic optimization was obtained using the Perdew-Burke-Ernzerhof~\cite{PhysRevLett.77.3865} exchange-correlation functional. We then apply a downfolding method to generate, from the Kohn-Sham Bloch functions, a set of symmetry-adapted Wannier functions~\cite{PhysRevB.56.12847,*Mostofi2008685} that can accurately describe the states around the Fermi level. In our case, these functions are boron and nitrogen $s$, $p$, and $d$ orbitals, and as can be seen in Fig.\,\ref{f.2}(B), the obtained 18-band spin-restricted $H_0$ accurately reproduces the DFT band structure in the energy interval of interest.  Both band structures also show a direct band gap $E_g=4.90$\,eV at the Brillouin zone corner points, in good agreement with experiments~\cite{doi:10.1021/nn1006495}.

\begin{figure}
	\includegraphics[trim = 0mm 0mm 0mm 0mm,scale=0.5,keepaspectratio,clip=true]{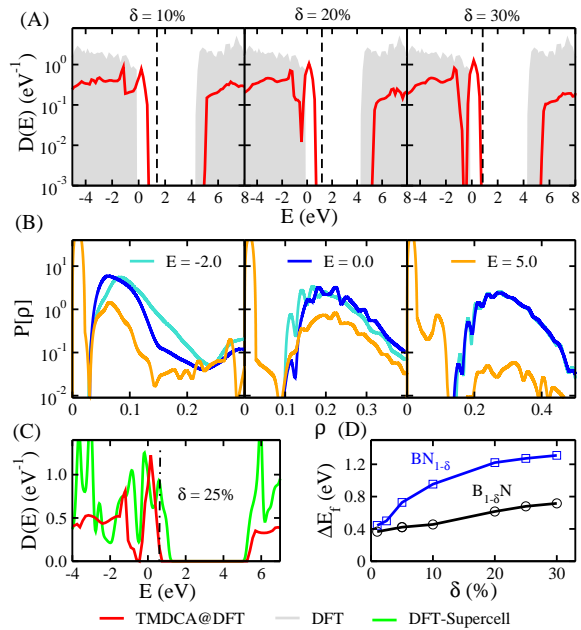}
	\caption{TMDCA@DFT results of $h$-BN obtained for $N_c=8$ and $U=0$.  (A) and (B) show the TDoS and the local density of states distribution, respectively, at three different $\delta$.  For reference, (A) also includes the density of states of pristine $h$-BN obtained from DFT and the Fermi level (vertical dashed lines) from the TMDCA SCF calculations.  (C) A comparison of the density of states obtained using the supercell approach with the TMDCA@DFT for $\delta=25\%$.  (D) The formation energy as a function of boron and nitrogen vacancy concentrations.}
	\label{f.3}
\end{figure}

Disorder can take many forms, including impurities, adatoms, and vacancies. Regardless of the origin, we describe disorder through the difference between the single-particle Hamiltonian for the disordered and pristine structures. Although this difference Hamiltonian comprises off-diagonal as well as diagonal elements, the former are generally smaller and for clarity has been dropped from Eq.\,(\ref{e.1}). In the boron nitride example, we focus on boron vacancies, as the B sublattice is more prone to defects due to the lower threshold energy for knock-on damage than the N sublattice~\cite{PhysRevB.75.245402,PhysRevB.80.155425,PhysRevLett.102.195505}. We confirmed this observation from our calculations of the vacancy formation energies in the B and N sublattices [see Fig.\,\ref{f.3}(D)].

We use a disorder potential comprised of binary site potentials $V_{i\sigma}^\alpha=V_i\in\{0,W\}$, where the two elements represent the absence and presence of a vacancy, respectively, with the vacancy potential $W$ being a lot greater than the material bandwidth.  Random disorder configurations are then generated using the probability mass functions $P(V_i=W)=\delta$ and $P(V_i=W)=0$ for B and N sites, respectively, where the averaged boron vacancy concentration $\delta$ satisfies the stoichiometry B$_{1-\delta}$N of the disordered material.

Electronic properties, including many physical measurable attributes, can be obtained knowing the single-particle Green function. Obtaining the full single-particle Green function for an infinite disordered lattice, however, is not feasible.  Therefore, we instead rely on approximate Green functions, in our case, based on the TMDCA.  The TMDCA is based on a formalism consistent with the generalized dynamical cluster theory approaches to correlated electron systems.  We refer interested readers to Refs.~\cite{RevModPhys.77.1027,Lee1985,*RevModPhys.66.261}.  The main steps of the TMDCA@DFT self-consistency are outlined below with additional details provided in SM~\cite{Supp}.

Through a set of SCF equations, the TMDCA approach maps a disordered lattice onto a finite cluster embedded in a typical medium, as illustrated in Fig.\,\ref{f.1}(A).  The cluster is a periodically repeated cell containing $N_c$ primitive cells, which results in the first Brillouin zone of the original lattice being divided into $N_c$ non-overlapping cells, where each cell centered at the wave vector $\K$ contains a set of wave vectors $\tildek\equiv\k-\K$, where $\tildek$ and $\k$ are wave vectors generated by the translational symmetry of the cluster and the original lattice, respectively~\cite{PhysRevB.63.125102,*PhysRevB.61.12739}.  The clusters allow for resonance effects and, by increasing $N_c$, we can systematically incorporate longer-range spatial fluctuations.  We recommend using a cluster lattice that preserves the symmetry of the original lattice, and in that vein, we adopt a hexagonal cluster lattice in our $h$-BN calculations with cluster lattice vectors being a multiple of the original lattice vectors shown in Fig.\,\ref{f.2}(A).

The self-consistency procedure in Fig.\,\ref{f.1}(A) goes as follows:  (i) We make an initial guess of a hybridization function $\Gamma(\K)$, which describes the coupling between the cluster and the effective medium.  (ii) We calculate a fully dressed cluster Green function $G^c(E)=(\mathcal G^{-1}-V-\Sigma^\mathrm{Int})^{-1}$, where $\mathcal G^{-1}$ is the cluster-excluded Green function, $V$ is the disorder potential, and $\Sigma^\mathrm{Int}$ is the second-order expansion of the electron interactions.  The self-energy $\Sigma^\mathrm{Int}$ is obtained self-consistently, as illustrated in the secondary SCF loop in Fig.\,\ref{f.1}(A).  (iii) We calculate the cluster density of states $\rho^c=-\frac{1}{\pi} \Im G^c$ and average over a large number of configurations to obtain the wave-vector-resolved, non-self-averaged typical DoS~\cite{PhysRevB.89.081107,PhysRevB.92.201114}
\begin{equation}
	\rho_{t}^c(\K) = \langle \rho^c_i\rangle_\mathrm{geom} \left\langle \frac{\rho^c(\K)}{\frac{1}{N_c} \sum_{i}\rho_{i}^c}\right\rangle_\mathrm{arit},
	\label{e.2}
\end{equation}
where $\langle \rho^c_i\rangle_\mathrm{geom}=\exp\,\langle\ln\rho_i\rangle_\mathrm{arit}$ is the diagonal elements of $\langle \rho^c\rangle_\mathrm{geom}$.  The purpose of the second factor is to capture non-local fluctuations.  (iv) A cluster typical Green function $G_t^c(\K)$ is then calculated from the Kramer-Kronig transform of Eq.\,\ref{e.2}, which is subsequently used to calculate the coarse-grained Green function
\begin{equation}
	\bar{G} (\K) = \frac{N_c}{N} \sum_{\tildek}  \bigg [\displaystyle G^c_{t} (\K)^{-1} + \Gamma (\K) - H_0(\k) + \bar{H}_0(\K) + \mu \bigg ]^{-1}
	\label{e.3}
\end{equation}
where the overbar depicts cluster coarse-graining and $\mu$ is the Fermi level, which we obtained in the secondary SCF loop mentioned in step (ii).  (v) A new hybridization function is obtained from
\begin{equation}
	\Gamma_\mathrm{n}(\K) = (1-\zeta)\Gamma_\mathrm{o}(\K)+\zeta\left[(G^c)^{-1}-\bar G^{-1}\right],
	\label{e.4}
\end{equation}
where $\Gamma_\mathrm{n}$ ($\Gamma_\mathrm{o}$) refers to the new (old) hybridization function and $\zeta$ is a mixing parameter.  $\Gamma(\K)\equiv\Gamma_\mathrm{n}(\K)$ is then used in (ii) to close the primary SCF loop.  Convergence is achieved when $G_t^c=\bar{G} $.  The convergence of the TMDCA@DFT formalism as function of increasing cluster size is discussed in the SM~\cite{Supp}.

Let us at this point focus on our $h$-BN example. Before proceeding to the general case, let us first consider the noninteracting limit.  Figure~\ref{f.3}(A) shows the typical density of states (TDoS) for an $N_c=8$ cluster at various B vacancy concentrations $\delta$.  For delocalized states, the TDoS is expected to be similar to the arithmetically averaged density of states (ADoS).  Close to the localization transition, however, the TDoS, unlike the ADoS, is a strongly varying, non-self-averaging quantity with substantial weight only on a few lattice sites~\cite{nakayama2003fractal,*Janssen1998,*Thouless1974,*Thouless1970,PhysRevB.89.081107,0953-8984-26-27-274209,Vlad2003}.  As $\delta$ increases in Fig.\,\ref{f.3}(A), we see structure develop at the valance band edge.  The Fermi level also shifts to lower energy defined with respect to $\delta=0$, implying $p$-type doping.  Even with a concentration as high as $\delta=30$\%, we find that the insulating phase is stabilized.  To further verify this, we performed relaxed supercell calculations at $\delta=6.2$\% (not shown) and $\delta=25$\% shown in Fig.\,\ref{f.3}(C).  Both calculations show insulating behavior, suggesting that vacancies alone are not sufficient to induce an IMT in $h$-BN.

To gain insight into the energetics of the vacancy formation and the stability of the disordered $h$-BN, we calculated the formation energy $\Delta E_f \equiv E_{v} - E_{0}$, where $E_{v}$ and $E_{0}$ are the total energies of the vacancy and pristine structures, respectively. As shown in Fig.\,\ref{f.3}(D), $\Delta E_f$ is positive and increases with increasing vacancy concentration, though the energy cost is less for B vacancies.  Additionally, the lack of a kink in the $\Delta E_f$ curves indicates an absence of a transition, a further confirmation that the insulating state is stabilized against disorder.

\begin{figure}
	\includegraphics[trim = 0mm 0mm 0mm 0mm,scale=0.5,keepaspectratio,clip=true]{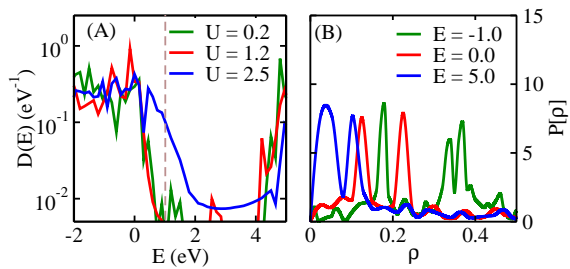}
	\caption{TMDCA@DFT results of $h$-BN obtained for $N_c=8$ and $\delta=1.0\%$.  (A) The TDoS for various $U$.  At small $U$, impurity states develop near the valence band edge and as $U=2.5$\,eV, these impurity states have hybridized with conduction band states, resulting in an insulator-metal transition.  The vertical dashed line depicts the Fermi level.  (B) The local density of states distribution at $U = 2.5$\,eV and various energies.  The distributions are characteristic of a disordered metal.}
	\label{f.4}
\end{figure}
Next, we investigate the combined effects of vacancies and electron interactions, focusing on the paramagnetic phase.  An exploration of the ($\delta$, $U$) parameter space revealed that an IMT occurs roughly at $\delta\times U^4\approx 0.8$\,eV$^4$, as shown in Fig.\,\ref{f.1}(C).  In Fig.\,\ref{f.4}(A), we show the TDoS for an $N_c=4$ cluster at $\delta=1.0\%$ and various $U$.  As in the noninteracting limit, impurity states develop within the gap for small $U$.  Around $U\approx2.5$\,eV, however, valence and conduction states merge, signaling a transition from a band insulator to a correlated metal.  Thus, the presence of interactions induces an IMT, even at a modest boron vacancy concentration. We attribute the IMT to the injection of mobile carriers caused by inelastic scattering processes hybridizing localized and delocalized states~\cite{10.1139,*Mott,*Mott1967,*Mott1949}. In our case, the hybridization occurs between impurity states from the valence and low-energy conduction band states.  There is no need for thermal activation; rather, the necessary energy is mostly provided by the electron interaction.  That these correlated impurity states delocalize~\cite{Luque2012,*Luque2006320} also means that the carriers do not decay into other localized states existing within the gap~\cite{PhysRev.87.835,*PhysRev.87.387}.  Instead, the disorder is screened~\cite{Ma1982,*PhysRevLett.91.066603,*Aguiar2013} in these hybridized states allowing them to become extended~\cite{10.1139,*Mott,*Mott1967,*Mott1949}.  This rich many-body physics underscores that electron interaction is critical in the description of the IMT in $h$-BN.

An important physical observable for characterizing disordered materials is the local density of states distribution P$[\rho]$, which could be measured in optoelectronic experiments.  More specifically, P$[\rho]$ is a well-defined fingerprint of the spatial variations of the local density of states, which tends towards Gaussian and log-normal distributions in the metallic and insulating phases, respectively~\cite{Anthony2010,PhysRevB.81.155106,PhysRevB.89.081107,PhysRevB.92.201114}, as illustrated in Fig.\,\ref{f.1}(B).  Figures~\ref{f.3}(B) and \ref{f.4}(B) show the normalized P$[\rho]$ calculated at various energies for various $\delta$ at $U=0$ eV and $\delta=1.0\%$ for $U = 2.7$\,eV, respectively. In the noninteracting limit, Gaussian distributions are observed for $E=-2.0$\,eV and $E=0$ but not for $E=5.0$\,eV.  This is expected as the latter is located in the gap below the conduction band, where the electrons are prone to being localized.  With interaction, P$[\rho]$ is reminiscent of a disordered metal for all the energies considered~\cite{Jaramillo2014,*Semmler2010,*Ruhlander200132}. Except for the locations of the peaks, note that P$[\rho]$ is qualitatively similar for the three energies.  We are not aware of any other computational studies of P$[\rho]$ in the presence of interaction.  In the interacting case, P$[\rho]$ could be identified as depicting the spatial nature of quasiparticle many-body excitations rather than single-particle states.  We speculate that the correlation length of these excitations, especially near the IMT, will be reduced due to inelastic processes and multisite scattering, including those from deeply trapped states.

In summary, we have presented a first-principles-based many-body approach for characterizing localization in disordered materials and applied it to study monolayer $h$-BN in the presence of electron interactions and randomly distributed boron vacancies.  Our calculations show an IMT, in which electron interactions play a critical role by hybridizing impurity states with low-energy states in the conduction band to form degenerate states within the gap.  This IMT opens up the possibility of conduction in $h$-BN.

This work has been funded by the Office of Naval Research, directly and through the Naval Research Laboratory.  CEE acknowledges the NRC Research Associateship Programs and XSEDE through NSF grant TG-DMR160027.  VD acknowledges NSF grant DMR-1410132. 

\textit{Note added --} We recently became aware of a related study of real materials in the noninteracting limit~\cite{PhysRevB.94.224208}. 
\ifx\mcitethebibliography\mciteundefinedmacro
\PackageError{unsrtM.bst}{mciteplus.sty has not been loaded}
{This bibstyle requires the use of the mciteplus package.}\fi

\setcounter{figure}{0}
\begin{widetext}
\section*{Supplemental Material on First-principles-based method for electron localization: Application to monolayer hexagonal boron nitride}
\section{Details of the TMDCA@DFT Self-Consistency}

To study the interplay of disorder and electron interactions, we utilize the Anderson-Hubbard Hamiltonian of the form
\begin{equation}
    H = H_0+\sum_{i\alpha\sigma} V_{i\sigma}^\alpha n_{i\sigma}^\alpha + U\sum_{i\alpha} n^\alpha_{i \uparrow} n^\alpha_{i \downarrow},
    \label{e.1}    
\end{equation} 
where $n_{i\sigma}^\alpha$ is the number operator, and $i$, $\alpha$, and $\sigma$ are site, orbital, and spin indices, respectively. The first term $H_0$ describes the single-particle Hamiltonian. This is obtained by downfolding the Kohn-Sham Bloch functions using a set of symmetry-adapted Wannier functions~\cite{PhysRevB.56.12847,*Mostofi2008685} that can accurately describe the states around the Fermi level. In our case, these functions are boron and nitrogen $s$, $p$, and $d$ orbitals of 18-band spin-restricted $H_0$ that accurately describes the DFT band structure in the energy interval of interest. The second term represents the disorder modeled by a binary site potentials $V_{i\sigma}^\alpha=V_i\in\{0,W\}$, where 0($W$) represent the absence (presence) of a vacancy, with the vacancy potential $W$ being a lot greater than the material bandwidth.  The vacancy is generated by a random disorder configurations using the probability mass functions $P(V_i=W)=\delta$ and $P(V_i=W)=0$ for B and N sites, respectively, where the averaged boron vacancy concentration $\delta$ satisfies the stoichiometry B$_{1-\delta}$N of the disordered material. The last term describes the Coulomb repulsion $U$ between two electrons occupying site $i$ incorporated self-consistently into our formalism 
via interacting, non-local cluster self-energy ($\Sigma_c[{ \tilde{G}}](i,j\neq i)$), which is up to second order in the perturbation expansion of the interactions, $U^2$.

As explained in the main text, we aim to obtain the single-particle Green function and the associated density of states (DoS). Obtaining these ``exactly'' for Eq.~\ref{e.1} is, however, not feasible. We instead, obtain approximate Green functions based on the typical medium dynamical cluster approximation (TMDCA). The TMDCA is a mean-field approach with an intrinsic order parameter based on the typical DoS that ``properly'' characterizes disordered and/or interacting electron systems even in the proximity of electron localization. The TMDCA self-consistency is based on a formalism consistent with the generalized dynamical cluster theory approaches to correlated electron systems~\cite{PhysRevB.63.125102,*PhysRevB.61.12739,*Jarrell01}. Through a set of self-consistency equations, the TMDCA approach maps a disordered lattice onto a finite cluster embedded in a typical medium, as illustrated in Fig.1(A) of the main text and explained therein, and described in details below. 

\begin{wrapfigure}{l}{0.2\textwidth}
\center{\includegraphics[trim = 0mm 5mm 5mm 15mm,width=.2\textwidth]{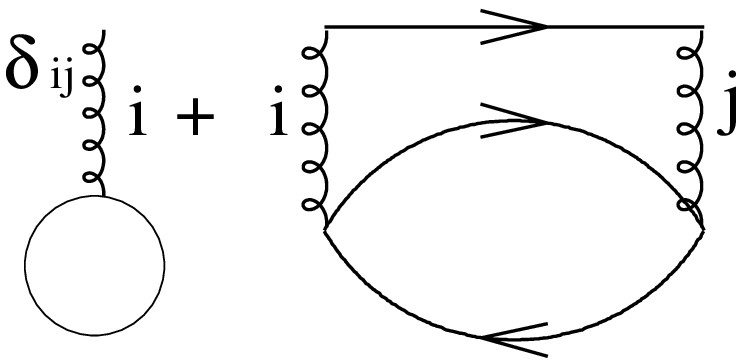}}
\caption{The first and second-order Feynman diagrams of the interacting self-energy between sites $i$ and $j$.}    
\label{Fig:sopt}     
\end{wrapfigure}

The self-consistency procedure follows:  (i) After obtaining $H_0$ from a converged, density functional theory (DFT) calculations by downfolding the Kohn-Sham Bloch functions using a set of symmetry-adapted Wannier functions~\cite{PhysRevB.56.12847,*Mostofi2008685}, we make an initial guess of a hybridization function $\Gamma(\K)$. The hybridization function measures the escape rate of electrons from the cluster to the effective medium. (ii) We Fourier transform the hybridization function to real space, $\Gamma_{n,m} = \sum_{\K} \Gamma(\K)\exp[i \K\cdot(\mathbf{R}_n-\mathbf{R}_m)]$ and then for a given vacancy configuration, form the cluster Green function $G^c(V)=(E-\Gamma -H_0-V)^{-1}$. (iii) Using $G^c(V)$, we construct a Hartree-corrected cluster Green function 
${\cal \tilde{G}}^{-1}_c(V,U) = G^c(V)^{-1} + \epsilon_d(U)$, where $\epsilon_d(U)=\mu -Un_i/2$ and $n_i$ $=$ $-1/\pi \int_{-\infty}^0  \Im {\cal \tilde{G}}(i,i,E) d E$ is the site occupancy at zero temperature. An additional internal self-consistency is used to converge $n_i$ and ${\cal \tilde{G}}_c$. This internal self-consistency ensures that $n_i$ is numerically the same as the one that could be obtained using the fully dressed Green function, but, at a fraction of the computation cost~\cite{PhysRevB.92.201114}. This is important as it is the most computational intensive part of the self-consistency. Also, it enables the systematic incorporation of crossing diagrams (for $N_c>1$) from both the disorder due to vacancy and electron interactions at equal footing~\cite{PhysRevB.92.201114}. (iv) For a given electron interaction strength and randomly generated disorder due to boron vacancy configuration $V$, the fully dressed cluster Green function $G^c(V,U)=(E-\Gamma -H_0 -V-\Sigma^\mathrm{Int})^{-1}$ is calculated, where $\Sigma^\mathrm{Int}$ is the second-order expansion of the electron interactions obtained self-consistently, as illustrated in the secondary loop in Fig.1(A) of the main text using the diagram shown in Fig.~\ref{Fig:sopt} herein. (v) With the fully dressed Green function, we calculate the cluster density of states $\rho^c=-\frac{1}{\pi} \Im G^c$ and average over many configurations to obtain the momentum-resolved, non-self-averaged typical DoS~\cite{PhysRevB.89.081107,PhysRevB.92.201114}
\begin{equation}
    \rho_{t}^c(\K) = \langle \rho^c_i\rangle_\mathrm{geom} \left\langle \frac{\rho^c(\K)}{\frac{1}{N_c} \sum_{i}\rho_{i}^c}\right\rangle_\mathrm{arit},
    \label{e.2}
\end{equation}
where $\langle \rho^c_i\rangle_\mathrm{geom}=\exp\,\langle\ln\rho_i\rangle_\mathrm{arit}$. The second factor ensures that non-local fluctuations are captured while the system remains in a typical medium.  (vi) A cluster typical Green function $G_t^c(\K)$ is then obtained from the Kramer-Kronig transform of Eq.\,\ref{e.2} as $ G_{t}^c(\K,E)=\int d E' \displaystyle \rho_{typ}^c(\K,E')/(E - E')$ and used to calculate the coarse-grained Green function
\begin{equation}
    \bar{G} (\K) = \frac{N_c}{N} \sum_{\tildek}  \bigg [\displaystyle G^c_{t} (\K)^{-1} + \Gamma (\K) - H_0(\k) + \bar{H}_0(\K) + \mu \bigg ]^{-1}
    \label{e.3}
\end{equation}
where the overbar depicts cluster coarse-graining and $\mu$ is the Fermi level, which we obtained in step (iii).  (vii) A new hybridization function is obtained using linear mixing 
\begin{equation}
    \Gamma_\mathrm{n}(\K) = (1-\zeta)\Gamma_\mathrm{o}(\K)+\zeta\left[(G^c)^{-1}-\bar G^{-1}\right],
    \label{e.4}
\end{equation}
where $\Gamma_\mathrm{n}$ ($\Gamma_\mathrm{o}$) refers to the new (old) hybridization function and $\zeta$ is a mixing parameter.  (viii) $\Gamma(\K)\equiv\Gamma_\mathrm{n}(\K)$ is then used in (ii) to repeat the above procedure until the hybridization function converges to the desired accuracy. When this happens, $G_t^c\equiv\bar{G} $ within the computational error.

\section{Convergence of the TMDCA@DFT with Cluster Size}
We show in Fig.\,\ref{f.2} the typical density of states (TDoS) for various cluster sites $N_c=1$, 4, 8, and 16.  Figure~\ref{f.2}(A) depicts the TDoS in the noninteracting limit at the boron vacancy concentrations $\delta=5$ and $10\%$, while Fig.\,\ref{f.2}(B) shows the the TDoS at the boron vacancy concentrations $\delta=0.5$ and $1.0\%$ at the electron interaction strength $U=1.2$ eV.  
\begin{figure}[b!]
    \includegraphics[trim = 0mm 0mm 0mm 0mm,scale=0.45,keepaspectratio,clip=true]{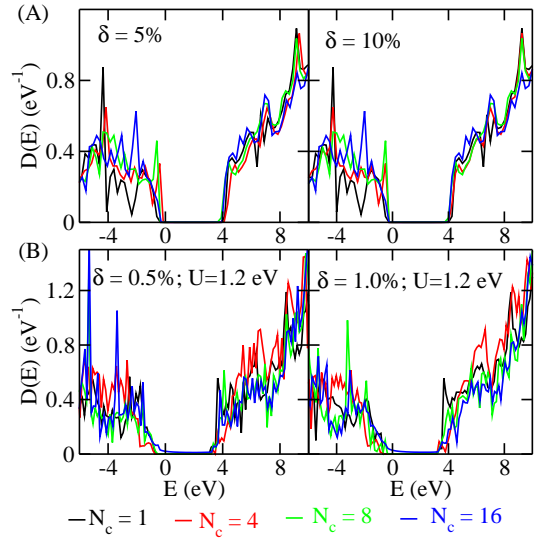}
    \caption{The TMDCA@DFT results of $h$-BN obtained for $N_c$ of 1, 4, 8, and 16 at two boron vacancy concentrations $\delta$ for (A) noninteracting and (B) interacting cases.  Note that the electrochemical potential has been shifted to allow direct comparison of the spectra.}
    \label{f.2}
\end{figure}
The essence of this plot is to benchmark the convergence of the TMDCA@DFT method.  In both cases, the TDoS systematically converges as $N_c$ increases throughout almost the entire energy spectrum. This rather fast convergence ensures that with relatively small $N_c$, accurate results can be obtained, enabling simulation of many-orbital material as in our hexagonal boron nitride example. Even the $N_c=1$ cluster gives qualitative results, which could be important for computations of larger systems.  Note also that in the interacting limit, there are additional scattering processes leading to the emergence of deep-level impurity states.   

This fast convergence also makes it difficult to carry out standard finite-size-scaling of the TDoS. However, since the TDoS profile at almost all energy is converged as a function of $N_c$, our calculations remain in the thermodynamic limit to within the computation error of our method. The TMDCA self-consistently calculates the crucial, critical order parameter describing the disorder-induced formation of electronic bound state, by focusing on the typical (most probable) local density of the states. The validity and even quantitative accuracy of the TMDCA applied to model systems have been established in a series papers. The TMDCA is benchmarked against exact numerical methods, e.g. Kernel polynomial method, Transfer matrix method, Exact diagonalization (see for, e.g., Refs.~\cite{PhysRevB.92.014209,*PhysRevB.90.094208,*PhysRevB.92.205111} for the Anderson model. Also, the TMDCA is benchmarked against the continuous time quantum Monte-Carlo simulations for interacting disordered electron systems using the Anderson-Hubbard model~\cite{PhysRevB.92.201114}.

\ifx\mcitethebibliography\mciteundefinedmacro
\PackageError{unsrtM.bst}{mciteplus.sty has not been loaded}
{This bibstyle requires the use of the mciteplus package.}\fi

\end{widetext}

\end{document}